\documentclass[preprint,amsmath,amssymb,superscriptaddress,showpacs]{revtex4}
\usepackage{graphicx}
\begin{document}
\renewcommand{\thesection}{\arabic{section}}
\renewcommand{\thesubsection}{\arabic{section}.\arabic{subsection}}
\renewcommand{\thefigure}{\arabic{figure}}
\baselineskip=0.5cm
\renewcommand{\thefigure}{\arabic{figure}}
\title{Formation of atomic nanoclusters on graphene sheets}

\author{M. Neek-Amal }
\affiliation{School of Physics, Institute for research in fundamental sciences, IPM 19395-5531 Tehran, Iran }
\affiliation { Department of physics, Shahid Rajaei University , Lavizan,
Tehran 16788, Iran}
\author{Reza Asgari~\footnote{Corresponding author: asgari@theory.ipm.ac.ir}}
\affiliation{School of Physics, Institute for research in fundamental sciences, IPM 19395-5531 Tehran, Iran }
\author{M. R. Rahimi Tabar}
\affiliation{Department of Physics, Sharif University of Technology,
11365-9161, Tehran, Iran}
\affiliation{Institute of Physics, Carl von Ossietzky University, D-26111
Oldenburg, Germany}

\begin{abstract}
The formation of atomic nanoclusters on suspended graphene sheets have been investigated
by employing a Molecular dynamics simulation at finite temperature. Our systematic
study is based on temperature dependent Molecular dynamics simulations of some transition and alkali atoms on suspended graphene sheets. We find that the transition atoms aggregate and make various size nanoclusters distributed randomly on graphene surface. We also report that most alkali atoms make one atomic layer on graphene sheets.
Interestingly, the potassium atoms almost deposit regularly on the surface at low temperature.
We expect from this behavior that the electrical conductivity of a suspended graphene doped by potassium atoms
would be much higher than the case doped by the other atoms at low temperature.
\end{abstract}
\maketitle

\section{Introduction}

Graphene is a newly realized two-dimensional electron system~\cite{novoselov,geim}
which has produced a great deal of interest because of the new physics which it exhibits and because of its potential as a new material for electronic technology. The agent responsible for many of
the interesting electronic properties of graphene sheets is the non-Bravais honeycomb-lattice
arrangement of carbon atoms, which leads to a gapless semiconductor
with valence and conduction $\pi$-bands.
States near the Fermi energy of a graphene sheet are described by a massless Dirac equation which have
chiral band states in which the honeycomb-sublattice pseudospin is aligned either parallel to
or opposite to the envelope function momentum.~\cite{polini} The linear energy-momentum
dispersion has been confirmed by recent observations~\cite{novoselov1,bostwich}.
There are significant efforts to grow
graphene epitaxially~\cite{berger} by thermal decomposition of Silicon Carbide
(SiC), deposition of graphene sheets on solid~\cite{banerjee}or by vapor deposition of hydrocarbons on catalytic metallic
surfaces which could later be etched away leaving graphene on an
insulating substrate.

An unusual feature of the single-atom-thick layer of carbon atoms
is the absence of strongly localization~\cite{meyer,ishigami} when charge impurities, the short range ripples and surface roughness exist on graphene sheets. The issue of localization in graphene has recently attracted some
attention and the chiral nature of electron behavior has been discussed
in the literature.~\cite{suzuura,mccann}
Suzuura and Ando~\cite{suzuura}
claimed that the quantum correction to the conductivity in graphene
can differ from what is observed in normal two-dimensional electron gas due to the nature of
elastic scattering in graphene. This is possibly because of changing the sign of
the localization correction and turn weak localization into weak
antilocalization for the region when intervalley scattering time is
much larger than the phase coherence time. Further consideration of
the behavior of the quantum correction to the conductivity in
graphene~\cite{mccann} conclude that this behavior is entirely
suppressed due to time-reversal symmetry breaking of electronic
states around each degenerate valley.

Another special feature of the graphene is the capacity for using as a gas sensor because of special characteristics of graphene.~\cite{schedin,wehling,gierz,meyer1,leenaerts} Graphene-based gas sensors enable sensitivity to detect individual  events when a gas molecule is absorbed by a graphene sheet. The absorbed molecules change the local carrier concentration in graphene. Constant mobility of charge carriers in graphene with increasing chemical doping of NO$_2$ has been observed in experiment.~\cite{schedin} Note that doping molecules add some charge carriers but also induces charged impurities. The later effect results decreasing of the mobility. Different possible scenarios have been discussed to prove the constant mobility of charge carriers.~\cite{schedin}

Chen {\it et al.}~\cite{chen} have reported a systematic theoretical and experimental studies of the charged impurities mechanism by monitoring a reduction of the charge carrier mobility. The density of the charged impurities is induced by adding potassium atoms onto a graphene surface placed on SiO$_2/$Si substrate in ultrahigh vacuum at low temperature. They have reported that the addition of charged impurities produces a more linear behavior of conductance, reduces the mobility and moreover indicated that the minimum conductance depends on charged impurities density.

In addition, the enhancement of
electrical conductivity in a single-wall carbon nanotubes which are doped by
alkali atoms has been studied both experimentally and
theoretically~\cite{RSLeenature,RSLeePRB,Gao}. Gao {\it et al.}~\cite{Gao} carried out a
Molecular dynamics simulation to find optimum structure of
the doped potassium atoms for different number of the potassium atoms in two different packing schemes.

A series of measurements on suspended
graphene have been performed by experimental groups \cite{Giem2008,lee} and showed that graphene has an extraordinary stiffness which can support an additional weight of many
crystalline copper nanoparticles. The results from STEM micrograph of
graphene sheets incorporated copper atoms showed an aggregation shape that nanoparticles with different sizes distributed on graphene randomly. On the other hand, the pinning of size-selected the gold and nickel atoms on graphite has been investigated~\cite{smith} by using the MD simulation and it is shown that the atoms aggregated on the surface and gold cluster are shown to be flatter and more spread out the nickel cluster which are more compact. Moreover, they calculated the pinning energy thresholds and showed that there is a good agreement with those measured in experiment.~\cite{praton}

Particle aggregation in materials science is a direct consequence of mutual attraction between particles, atoms or molecules via van der Waals forces or chemical bonding.~\cite{elimelech} When there are collisions between particles, there are chances that particles will attach to each other and become a larger particle. There are three major physical mechanisms to form aggregate; Brownian motion, fluid shear or motion and differential settling. Because of interparticle forces, not all collisions may be successful in producing aggregates. If there is strong repulsion between particles then practically no collision gives an aggregate. It is known that collisions bring out by Brownian motion do not generally lead to the rapid formation of very large aggregates. In addition, the differential settling becomes more important when the particles are large and dense and this mechanism can promoting aggregation.

In general, there are two main methods for investigating the physical behaviors of the absorbed particles on a surface. A Monte Carlo method which is a sophisticate approach and the other one is a Molecular dynamics simulation. The purpose of this paper is the study of deposition and aggregation of atoms on graphene sheets by using the Molecular dynamics simulations.  Eventually, we generally report that the potassium atoms arrange regularly and deposit on a suspended graphene sheet at low temperature. This is an evidence that the electrical conductivity of the doped potassium atoms  on a suspended graphene should be higher than the electrical conductivity of other examined atoms.

It is worthwhile to note that since adding atoms are placed in the setback distance from graphene sheet, the effect of charged impurities can be negligible for the suspended graphene due to large screening effects. There are at least two reasons to explain why the effect of charge impurities should be important in graphene placed on substrate. The first effect comes from the fact that the averaged background dielectric constant value of graphene placed on a substrate is more or less three times larger than the vacuum dielectric constant. The second effect is related to the image of charged impurity~\cite{khanh} induces in another side of the dielectric substrate when there is a charged impurity above graphene. In the latter case, an induced electric dipole in the system increases the effect of electron scattering. We consequently expect that the difference between the carrier mobility for a suspended graphene and a graphene placed on substrate incorporating both the potassium atoms at low temperature.

The contents of the paper are described briefly as follows. In Sec.\,II, we
introduce the models and theory. Section III contains our numerical calculations. Finally, we conclude in Sec.\,IV with a brief summary.

\section{Model and Theory}

We have used the empirical inter-atomic interaction potential, carbon-carbon interaction
in graphite~\cite{brenner}, which contains three-body
interaction for Molecular dynamics simulation of a suspended graphene sheet at finite
temperature. The two-body potential gives a description of the
formation of a chemical bond between two atoms. Moreover, the
three-body potential favors structures in which the angle between
two bonds is made by the same atom. Many-body effects of electron
system, in average, is considered in the Brenner potential~\cite{brenner}, through
the bond-order and furthermore, the potential depends on the local
environment.

We have considered a graphene sheet
including $280 \times 160$ atoms with periodic boundary
condition. The area of a graphene sheet is $A_0=1173.5$~nm$^2$. Considering the canonical ensemble (NVT), we have
employed Nos\'{e}-Hoover thermostat to control temperature. Our
simulation time step is $1~fs$ in all cases and the thermostat's
parameter is $5~fs$. Therefore, we have found a stable two-dimensional graphene
sheet in our simulation.

First of all, we have simulated the stable graphene sheet at finite temperature, and then extra atoms have been  distributed randomly on graphene. The number of adatoms is equal to $2500$ and distributed in a portion of area, $A=\frac{4}{9}A_0$ in order to avoid boundary effects. We let system achieves to its equilibrium condition where the atoms above the surface are mostly fluctuating around their equilibrium positions. At the beginning of simulation, the atoms are initially located on the height equal to $3.65\dot{A}$ above the graphene sheet. To further proceed, we do need implement the interparticle interactions. We have used van der Waals potentials
between atom-atom and carbon-atom denoted by $X-X$ and $C-X$, respectively. The parameters of two-component interactions between two types of atoms can be estimated by simple average expressions
proposed by Steel {\it et al,} \cite{Steel}
\begin{eqnarray}
\sigma_{C-X}&=&\frac{\sigma_{X-X}+\sigma_{C-C}}{2}\nonumber\\
\varepsilon_{C-X}&=&\sqrt{\varepsilon_{X-X}\cdot\varepsilon_{C-C}}~,
\end{eqnarray}
where $\sigma_{i-j}$ is the collision diameter and $\varepsilon_{i-j}$ is the depth of primary energy well between  $i$th and $j$th atoms. The parameters which have been used in our simulations are given in Table.\ref{table1} adopted from Ref.~[\onlinecite{erkoc2001}]. It is worthwhile to note that our final results are independent of the explicit values of parameters appearing in von der Waals potentials. We have used new set of parameters~\cite{guan,ulbricht,hu} and our final results did not change.

\begin{table}
\begin{tabular}{|c|c|c|c|c|c|}
 \hline
  element & $\sigma$ &$\varepsilon $ & Atomic mass   \\
    &(\AA)   &(eV)& (gr/mol)  \\
  \hline
  Cu & 2.338& 0.40933& 63.546 \\
  Ag & 2.644& 0.3447& 107.8682 \\
  Au & 2.637 & 0.44147 &196.96 \\
  Li & 2.839 & 0.2053 & 6.941 \\
  Na & 3.475 & 0.1378 & 22.98977\\
  K & 4.285 & 0.11444 & 39.09\\

  C & 3.369& 0.00263& 12.01 \\

  Cu-C&2.853 &0.0328&-\\
  Ag-C&3.006 &0.0301&-\\
  Au-C&3.003 &0.0341&-\\
  Li-C &3.104 &0.0232&-\\
  Na-C &3.422 &0.0190&-\\
  K-C &3.827 &0.0173&-\\

  \hline
\end{tabular}
  \centering
  \caption{Physical parameters and Lennard-Jones $12-6$ parameters for some atoms and molecules.}\label{table1}
\end{table}

\section{Numerical Results}
In this section we present our numerical results based on the method described above. To this purpose, we have considered a system incorporating the transition atoms like copper, silver and gold atoms or the alkali atoms like lithium, sodium and potassium atoms on graphene sheets. Note that the gravitational force is very small in comparison to interparticle forces and tends to go unnoticed. Our main results have been summarized in Fig.~1. The copper atoms (Fig.~1a) are aggregated and the potassium atoms are deposited (Fig.~1b) on graphene sheets at $T=50$K.

Here, a graphene sheet plays the role of a substrate for external absorbed
atoms. Fig.~2 shows three snapshots of distributed
nanoclusters on graphene at $T=50$K. The transition atoms show aggregation configurations with producing different nanoclusters size. Apparently each nanocluster trapped around the height of out-of-plane because of graphene roughness and preferably made a nanocluster. Furthermore, since $\varepsilon_{cu-C}/\varepsilon_{cu-cu}$ is smaller than $\varepsilon_{Au-C}/\varepsilon_{Au-Au}$, the copper atoms thus distribute laterally on the surface much more than the gold atoms. In case, the gold atoms prefer to grow more vertically in comparison to the copper atoms. Our numerical simulations confirm such behaviors. Furthermore, our finding for the copper atoms is quite similar to those observed from STEM micrograph of
graphene sheets incorporated copper atoms by experimental group.~\cite{Giem2008} To be sure about the independency of interparticle interaction potentials, we have examined the Morse potential~\cite{erkoc2001} and we entirely get the same results.

 We have seen, on the other hand, new physics by adding the alkali atoms on graphene sheets at $T=50$K which the results are depicted in Fig.~3. The lithium atoms form a few layers of nanoclusters on graphene apparently having regular shape. Importantly by using the sodium atoms, an atomic layer thick arranged above the surface. This layer shows a sort of percolation configuration. Surprisingly, the potassium atoms arranged regularly and noticeably wet the graphene's surface. There are some small vacant islands where the potassium atoms could not occupy because they are bounded locally by bumps on the graphene surface. For graphene the doping is usually realized by surface transfer doping.

 The electronic properties that result from absorption depend on the ionic and covalent character of the bonds formed between carbon and the metal atoms.
 As alkali atoms easily release their valance electron, they may effectively induce n-type doping.~\cite{bostwich} Since the alkali atoms play the role as a cation, it can give electrons to the system easily. The amount of charge transfer to graphene by potassium atoms is a challenge problem and has been studied for many years in potassium deposited graphite surface.~\cite{Caragiu} The charge transfer between a paramagnetic molecule $NO_2$ and a graphene layer has been recently calculated~\cite{peeters} and found charge transfer of $1$e per molecule. Furthermore, The effect of potassium doping on the electric properties of graphene by using density functional theory (DFT$+U$) has been recently studied.~\cite{uchoa} They predicted a charge transfer of $0.51$e per potassium atoms at $U=0$. As $U$ increases, the charge transfer is linearly suppressed.

 The regular arrangement distributions of potassium (K) atoms on top of the surface indicate
 that the value of opening gap due to breaking symmetry would be much smaller than those values induced by the other metallic adatoms. Moreover, It is shown~\cite{alireza} that the conductance of charge carriers decreases by increasing the value of gap. Therefore, the K atoms give rise to increase the electrical conductivity, however the transition atoms evidently play a role as the central scatterers and result in decreasing the electrical conductivity of system. Note that the effect of charged impurities from chemical doping are negligible for suspended graphene~\cite{wehling}. Our results regarding to the regular arrangement of K atoms on graphene are similar to the alkali atoms doped on the single wall carbon nanotube.~\cite{Gao}

It is well known~\cite{elimelech} that for the alkali metal cations, the critical coagulation concentration values which estimated by assuming that no energy barrier exists, decreases in the order Li$^+>$Na$^+>$K$^+$ shows that the most hydrated ion, (Li$^+$) is the least effective in reducing repulsion term in particle-particle interactions. Consequently, the K atoms deposit easily on the surface respect to Li and Na atoms.

 We expect physically that the atoms escape from graphene sheets when temperature increases because of increasing the kinetic energy of atoms. The number and sizes of nanoclusters decrease by increasing temperature up to $T=300$K and at higher temperature atoms escape from the surface. In Fig.~4, we have shown the aggregation behavior of transition atoms at room temperature. The number of nanoclusters decrease by increasing temperature. Similar to the results shown in Fig.~2, the nanoclusters have not any regular crystalline configuration at higher temperature and moreover the distribution of potassium atoms changes and they show a percolation on top of graphene sheets as shown in Fig.~5.

For estimating the typical size of nanoclusters at different temperature, we have calculated the density number of copper atoms, $P(N)$ as a function of $N$, where $N$ is
the number of copper atoms in a nanocluster as depicted in Fig.~6. The number of copper atoms in
each nanocluster decreases by increasing temperature and some of the atoms escape from the surface.

One of the interesting subject for studying alkali metals
absorption on the graphite is the phase transition in the
structure shapes of overlayer.~\cite{Caragiu}
Most studied alkali metal on the graphite is potassium. A common
convention for overlayer structure is $2\times2$ phase in which 8
carbon atoms are surrounded by one K atom \cite{Caragiu} which
is most condensed phase of alkali atom absorbtion on the graphite
sheet in temperature lower than room temperature. There are other
less condensed structure such as $7\times7$ phase where 98 carbon
atoms are surrounded by one K atom. Two relevant
parameters for change in the coverage type in the graphite sheet
are temperature and density of adatoms. Increasing the number of
K atoms on the graphite at low temperatures ($\sim 90 K$)
closed pack islands ($2\times2$) covers all the entire surfaces
but also it may be the existence of the other low condensed phase.
After saturation of K atoms over the graphene it starts to make
other layers over the first one.~\cite{Caragiu} In the case of Li
atom most observation indicates on the intercalation into the
graphite even at $100$ K~\cite{26}. For Na atoms growing some
multilayered non ordered islands over each other have both
experimentally and computationally been observed~\cite{38}.

In our calculation, quite interestingly, the atoms pattern ($2\times2$), one K atom per eight carbon, for overlayer in high
density for K atoms and a snapshot of the structure is shown in Fig.~7.
This pattern is similar to that predicted for K atoms on graphite. In the low density, on the other hand,  we did not observe low dense phases, ($7\times7$) whereas small islands of the $2\times2$ structure
have been observed. Absence of low dense structures might be understood
from the dominant of thermal vibration which is larger in a strictly two dimensional
system with respect to three dimensional structure in graphite.

Furthermore, the type of ordering of sodium atoms on the
$C_{60}$ molecules has been investigated by Roques \emph{et al}~ \cite{Roques}. Their results showed that the ordering of deposited atoms is temperature and concentration dependence. They
found that up to eight atoms on the surface of $C_{60}$ molecule
there is no homogenous deposition and they begin to form
nano-clusters on the surface. This is because of saturation of
charges transfer from alkali atom to the substrate when the
substrate has no enough surface ( limiting by curvature) for accepting other charges.
In comparison with our calculations, crudely deducing, a graphene sheet with the very large radius of carbon cage, has enough available surface for
adding more alkali atoms and those alkali atoms enable to wet the
surface more and more.

We are also interested in calculating the out-of-plan carbon atoms in the presence of external atoms above graphene. The common procedure for measuring the roughness
exponents of a rough surface is to use a surface structure
function~\cite{fasolino,nima}
\begin{equation}
S_x(\delta)=<|h(x+\delta,y)-h(x,y)|^2>~,
\end{equation}
where the average is taken over some different $y$ values. The variation of $S_x(\delta)$ for different systems at $T=50$ K are shown in Fig.~8. The characteristic length is defined by the position that curves are bended. From our finding, the characteristic length in all cases are less than the case of graphene without absorbed atoms. Moreover, the characteristic length of the system by using the transition atoms are smaller than when the alkalis atoms are used. The reason is as follow, the nanocluster made by transition atoms distribute randomly over the graphene sheets increase the surface randomness. The $S_x(\delta)$ refereing to the situation that the potassium atoms added on graphene is very similar to one which graphene is itself alone because correlation between the potassium atoms and the carbon atoms are strong at low temperature. Consequently, this sort of atoms added on graphene does not change the morphology of surface at low temperature.

Another interesting quantity is normal-normal correlation for a surface.
The normal-normal correlation in the $x-$ direction is defined as
\begin{equation}
c(\delta,T)=<\hat{n}(x,y).\hat{n}(x+\delta,y)>~.
\end{equation}
The persistence length, $l_p(T)$ which is a criterion of surface stiffness is expressed as follow\cite{safran}
\begin{equation}
l_p(T)=\frac{\delta}{\ln c(0,T)-\ln c(\delta,T)}~.
\end{equation}

A surface is rigid in a length smaller than persistence length and it behaves as a soft membrane for longer length. To calculate the adatom dependence of
persistence length of graphene sheets, we have calculated this parameter for different used atoms at
three temperatures which are listed in Table~\ref{table2}. Moreover, by using the Morse potential for the transition atoms, the same persistence length values have been obtained.

\begin{table}
\begin{tabular}{|c|c|c|c|}
 \hline
  element   &$l_p$(\AA) at T=50 K&$l_p$(\AA) at T=200 K  &$l_p$(\AA) at T=300 K \\
  \hline
  g {\&} Cu & 88.1&60.0 &52.0\\
  g {\&} Ag & 82.5&58.0&52.0\\
  g {\&} Au & 87.2&58.5 &52.1 \\
  g {\&} Li & 95.3 &60.0& 54.3\\
  g {\&} Na & 94.5& 60.5&53.5\\
  g {\&} K  & 102.1 & 67.2&54.0 \\
  g &112.0&69.0&55.5 \\

  \hline
\end{tabular}
  \centering
  \caption{Persistence length of graphene (denoted by g) and the presence of adatoms on graphene at different temperatures.}\label{table2}
\end{table}

As it is clear from the Table~\ref{table2}, the persistence length decreases by incorporating atoms and importantly the persistence length of the potassium atoms is close to $l_p$ of graphene itself. In all cases, $l_p$ decreases by increasing temperature.
\section{Conclusion}
We have studied the formation of atomic nanoclusters on suspended graphene sheets by using a Molecular dynamics simulation at finite temperature. We have used the model of interparticle potentials.
The Brenner potential for carbon-carbon interactions potential and van der Waals model potentials for the interactions between atoms and moreover the interactions between atoms and the carbon atoms. We have shown that the transition atoms aggregated on the surface with different nanoclusters sizes, however the sodium and the potassium atoms produced one atomic layer on graphene. Interestingly, the potassium atoms arranged regularly on graphene sheets at low temperature and it indicates that the value of opening gap due to breaking symmetry would be much smaller than those values induced by the other metallic adatoms.

As a consequence, the charge carriers electrical conductivity of a suspended graphene doped by the potassium atoms would be higher than a suspended graphene doped by the other atoms. We qualitatively expect that the electrical conductivity of system would increase by adding the potassium atoms on suspended graphene sheets at low temperature. Note that since the induced charged impurities are set far from the suspended graphene sheet, the effect of charged impurities can be ignored due to large screening effects. Our finding at low temperature would be verified by experiments.

We remark that a model going beyond
the Molecular dynamics simulation is necessary to account quantitatively the effect of the potassium atoms on electrical conductivity for a suspended graphene sheet at low temperature. One approach would be the density-functional theory together with the Molecular dynamics simulation where the transport properties of charge carriers in the presence of
extra atoms are considered.

\begin{acknowledgments}

We thank Andre Geim for pointing this problem out to us. We are grateful to N. Abedpour for discussions
and comments. R. A. would like to thank the International Center for Theoretical Physics, Trieste for its hospitality during the period when part of this work was carried out.
\end{acknowledgments}

\newpage
\begin{figure}[ht]
\begin{center}
\includegraphics[width=0.54\linewidth]{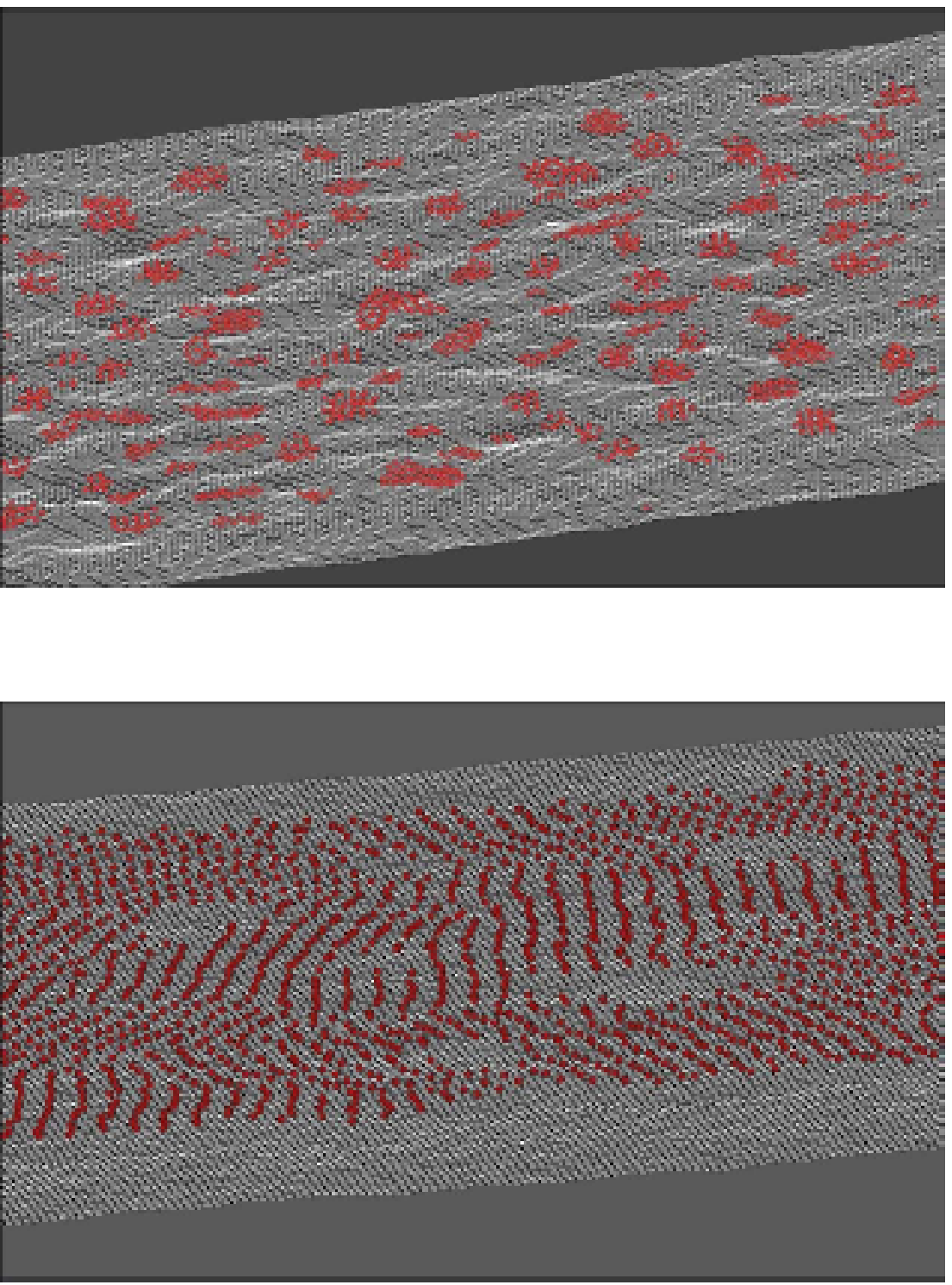}
\caption{(Color online) Distribution of copper atoms (top) and potassium atoms (bottom) on graphene sheets at $T=50$ K. The copper and potassium atoms denoted by red spheres and graphene sheets denoted by white spheres as a background. The number of atoms added on graphene is $2500$ for both cases. }
\end{center}
\end{figure}

\begin{figure}[ht]
\begin{center}
\includegraphics[width=0.52\linewidth]{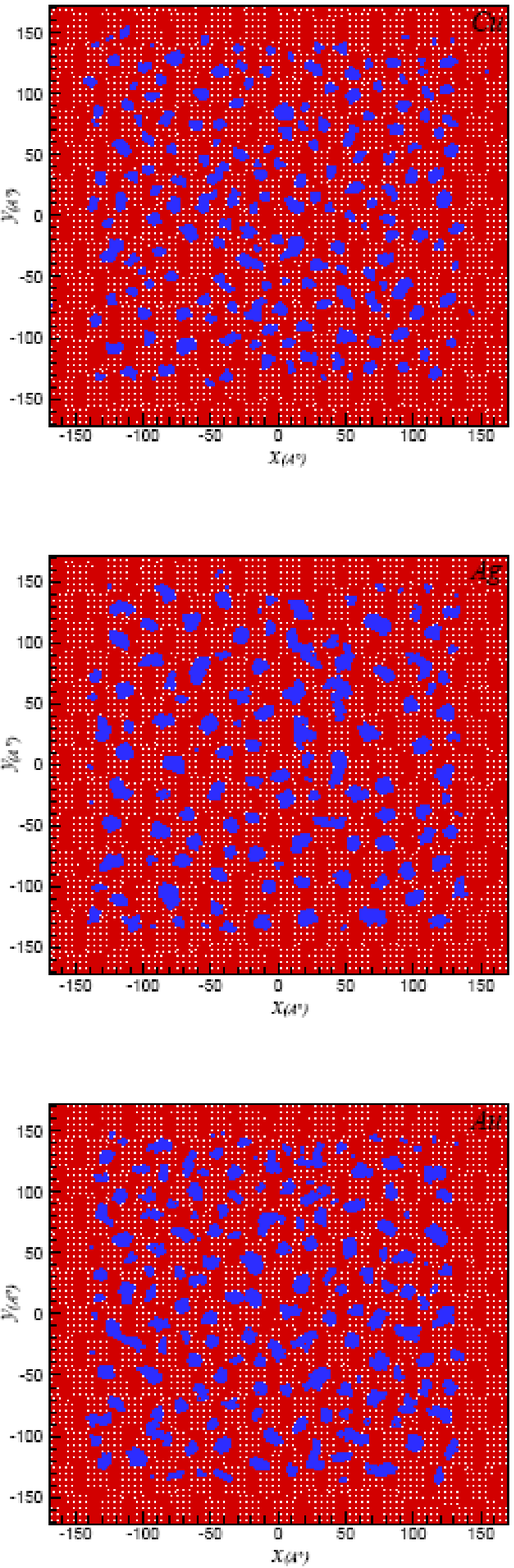}
\caption{(Color online) Aggregation of transition atoms on graphene sheets at $T=50$ K. Nanoclusters are distributed randomly above the surface. }
\end{center}
\end{figure}

\begin{figure}[ht]
\begin{center}
\includegraphics[width=0.52\linewidth]{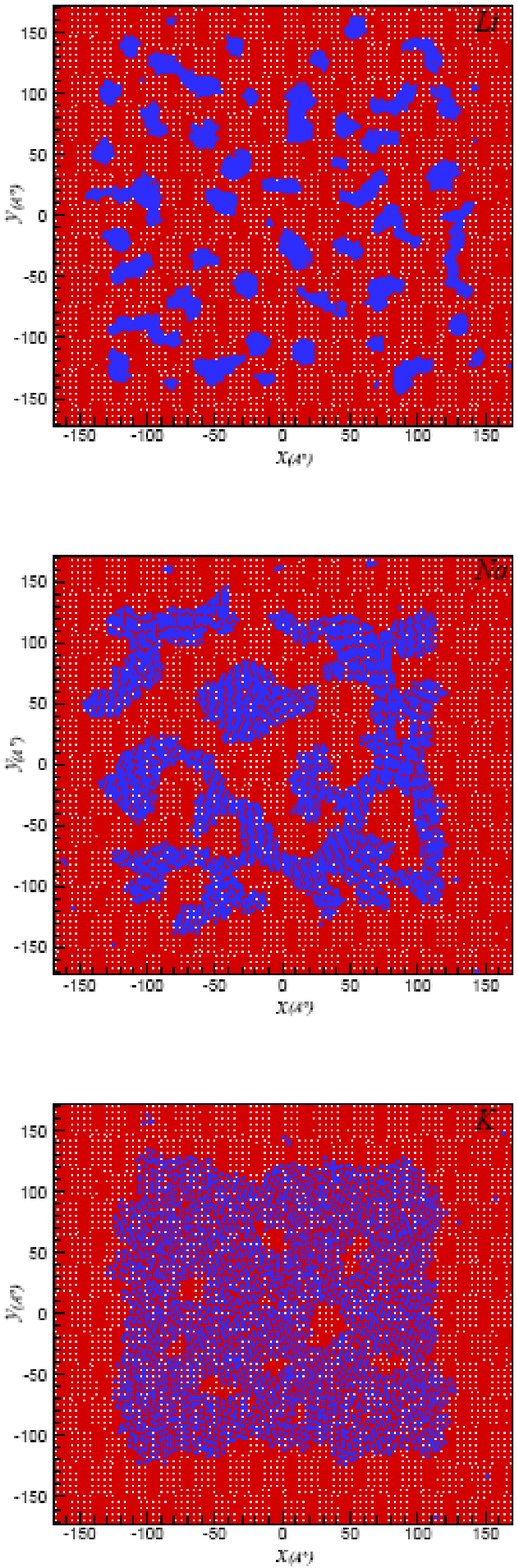}
\caption{(Color online) Distribution of alkali atoms on graphene sheets at $T=50$ K. Potassium atoms deposit on graphene. }
\end{center}
\end{figure}

\begin{figure}[ht]
\begin{center}
\includegraphics[width=0.52\linewidth]{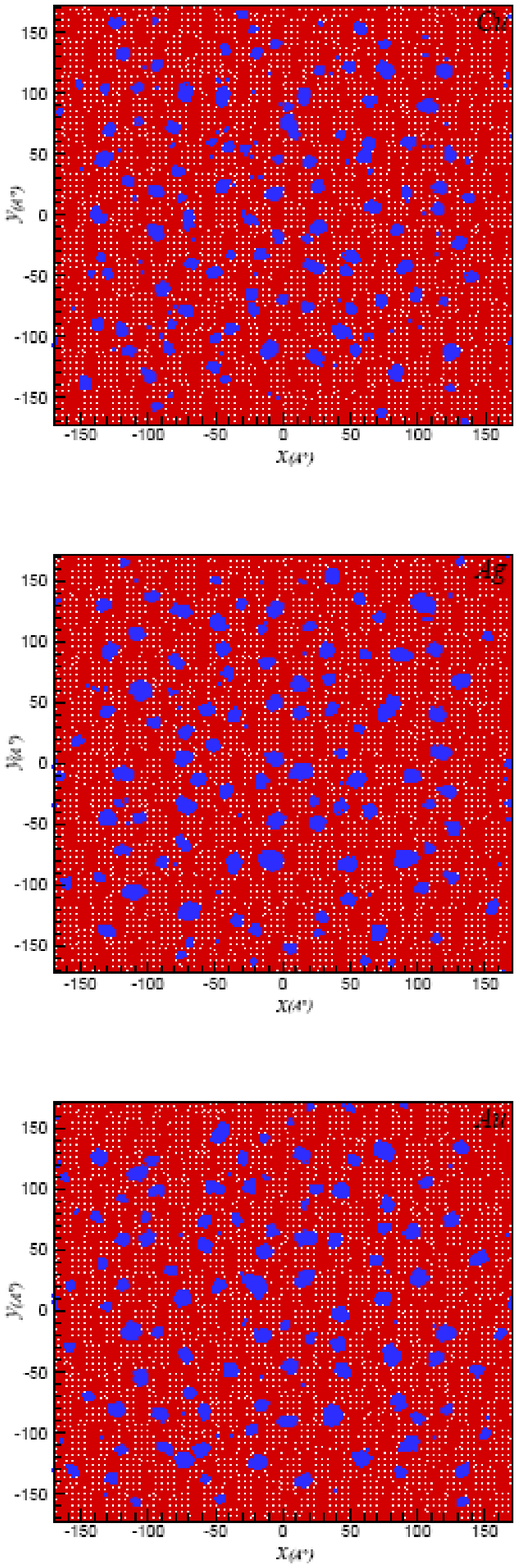}
\caption{(Color online) Aggregation of transition atoms on graphene sheets at $T=300$ K. }
\end{center}
\end{figure}

\begin{figure}[ht]
\begin{center}
\includegraphics[width=0.52\linewidth]{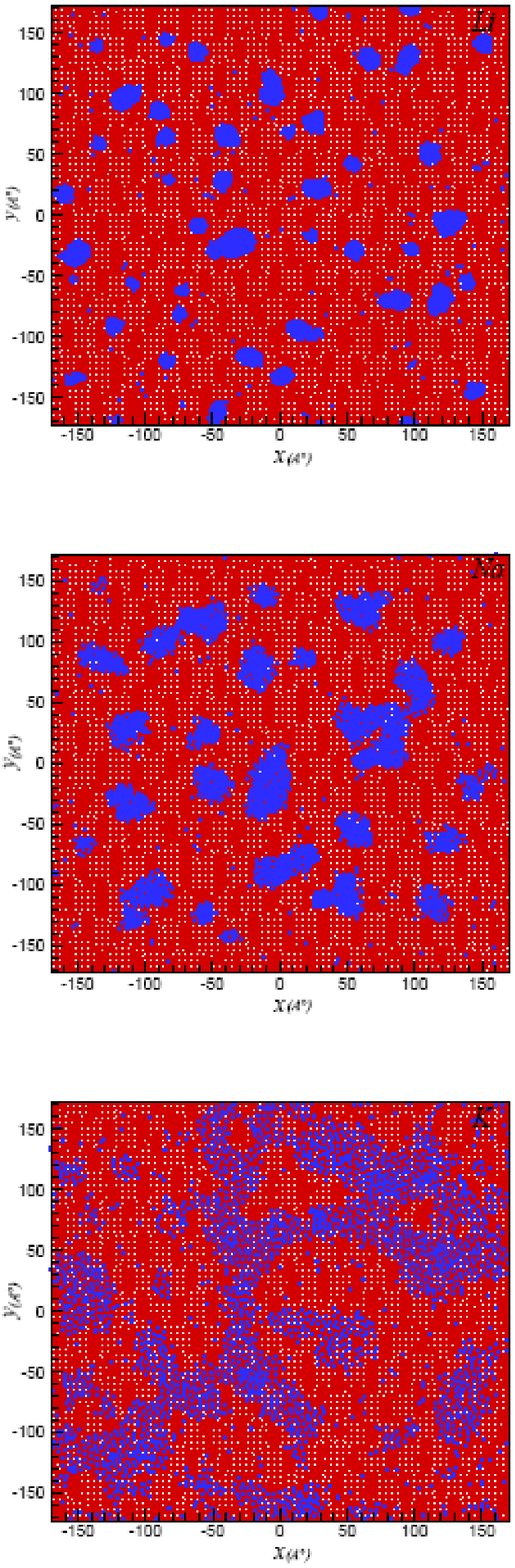}
\caption{(Color online) Distribution of alkali atoms on graphene sheets at $T=300$ K. }
\end{center}
\end{figure}

\begin{figure}[ht]
\begin{center}
\includegraphics[width=0.45\linewidth]{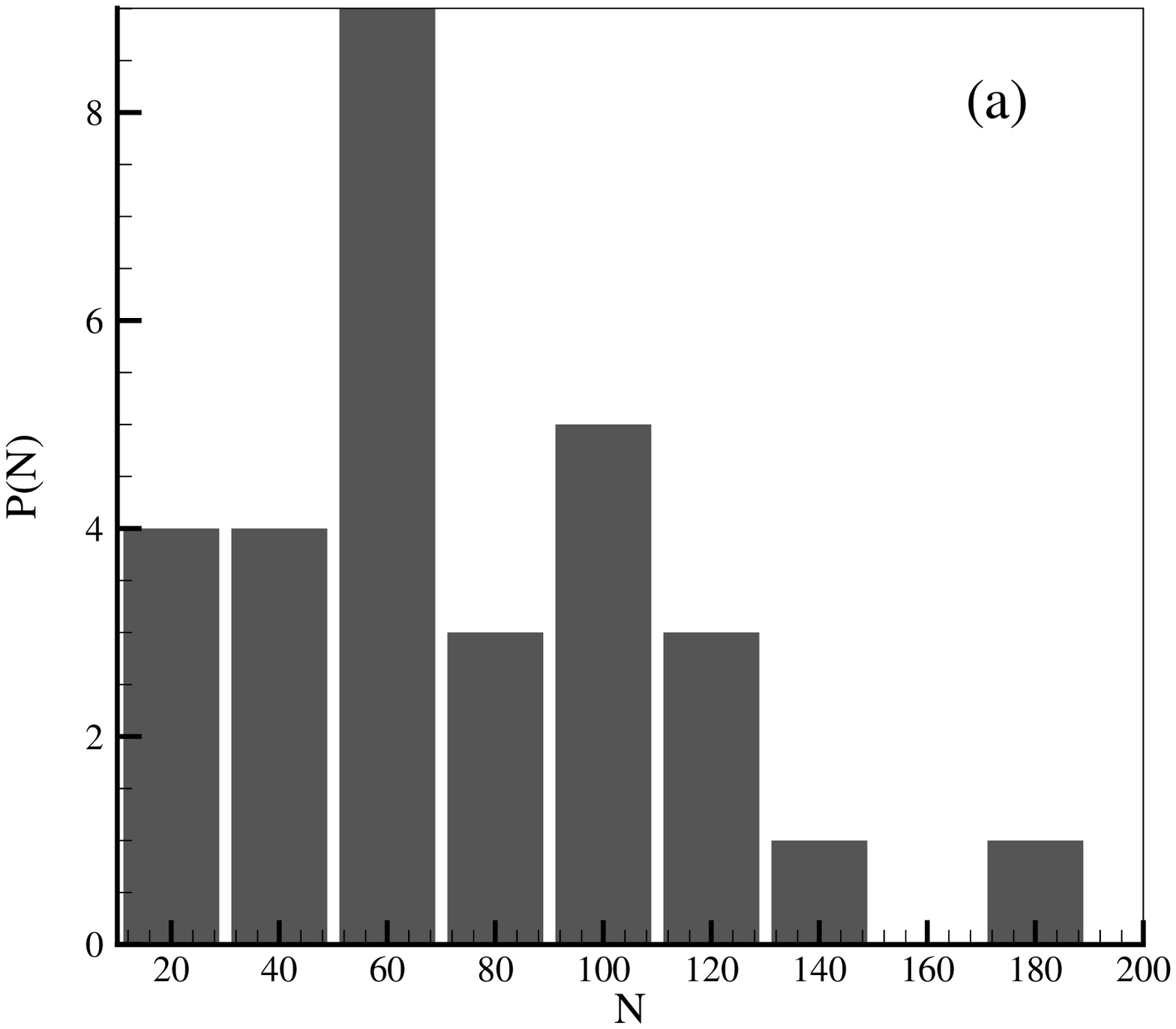}
\includegraphics[width=0.45\linewidth]{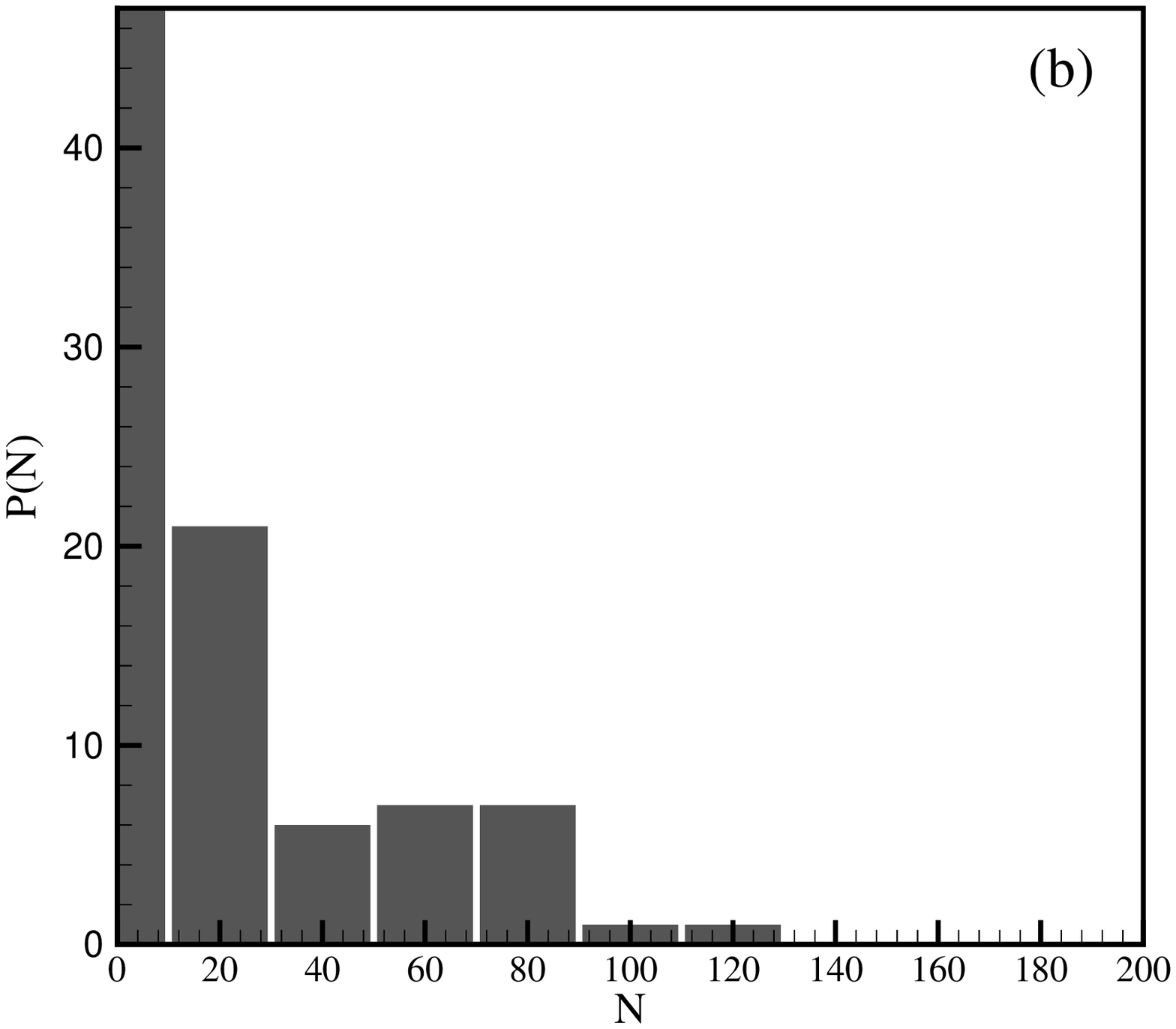}
\caption{(Color online) Density number of copper atoms as a function of number of atoms at $T=50$K (a) and $T=300$ K (b) .}
\end{center}
\end{figure}

\begin{figure}[ht]
\begin{center}
\includegraphics[width=0.52\linewidth]{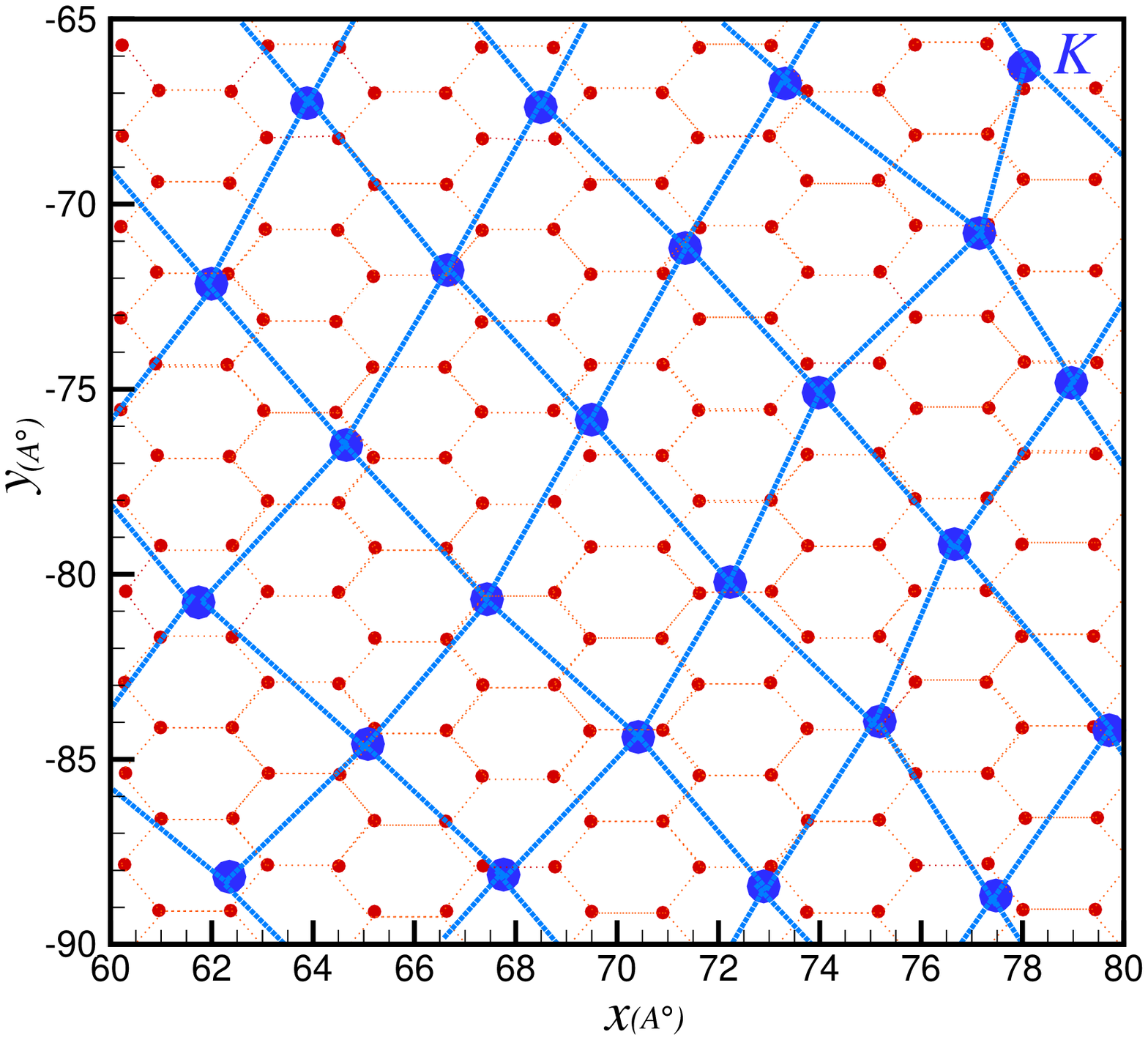}
\caption{(Color online) Pattern of potassium atoms distribution, (2 $\times$ 2) on a graphene sheet at $T=50$ K. The big circles represent the potassium atoms and the small dots represent the $C$ atoms.}
\end{center}
\end{figure}

\begin{figure}[ht]
\begin{center}
\includegraphics[width=0.52\linewidth]{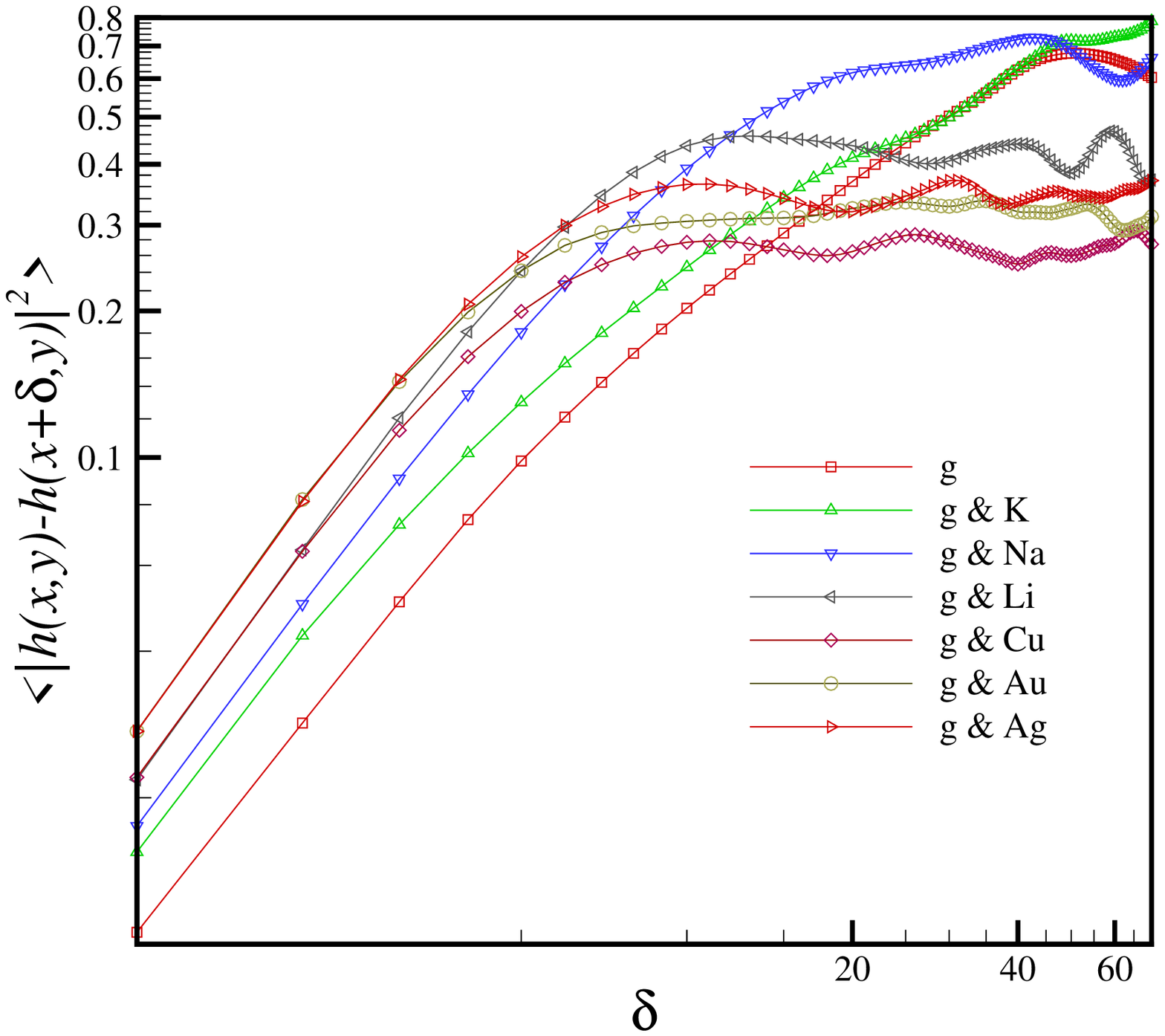}
\caption{(Color online) Surface structure function, $S_x(\delta)$ as a function of $\delta$ at $T=50$K. Graphene is denoted by $g$.}
\end{center}
\end{figure}

\end{document}